\documentclass[pra,a4paper,twoside,twocolumn,english,showpacs]{revtex4}
\pdfoutput=1
\usepackage[T1]{fontenc}
\usepackage[latin9]{inputenc}
\usepackage{textcomp}
\usepackage{color}
\usepackage{graphicx}
\usepackage{amssymb}

\makeatletter

\providecommand{\boldsymbol}[1]{\mbox{\boldmath $#1$}}

\usepackage{hyperref}

\usepackage{babel}
\makeatother

\begin{document}

\title{Non-ideal atom-light interfaces: modeling real-world effects}

\author{M. Koschorreck}

\email{marco.koschorreck@icfo.es}
\homepage{www.icfo.es}

\author{M. W. Mitchell}

\address{ICFO - Institut de Ciencies Fotoniques, Mediterranean Technology
Park, 08860 Castelldefels (Barcelona), Spain}

\begin{abstract}
We present a model which describes coherent and incoherent processes
in continuous-variable atom-light interfaces. We assume Gaussian states
for light and atoms and formulate the system dynamics in terms of
first and second moments of the angular momentum operators. Spatial
and temporal inhomogeneities in light and atom variables are incorporated
by partitioning the system into small homogeneous segments. Furthermore,
other experimental imperfections as for instance limited detector
time-resolution and atomic motion are simulated. The model is capable
of describing many experimental situations ranging from room temperature
vapor cells to sub-mK atomic clouds. To illustrate the method, we
calculate the effect of detector time-resolution, spatial inhomogeneities
and atomic motion on the spin squeezing dynamics of rubidium 87 on
the $D_{2}$ transition.
\end{abstract}

 \maketitle

\section{Introduction}

In the last decade, many pioneering experiments have demonstrated
quantum information processing with continuous variables in atomic
and photonic systems. Atomic spin squeezing can overcome the standard
quantum limit in magnetometry \citep{Wineland1992PRAv46pR6797}. Macroscopic
numbers of atoms can serve as a memory of a quantum state of light
\citep{Julsgaard2004Nv432p482}. A variety of atomic systems have
been proposed or demonstrated, including hot atoms in cells and cold
atoms in magneto-optical traps and in optical dipole traps. The conditions
vary greatly from system to system, notably in number of atoms, from
$\sim10^{12}$ in cells to $\sim10^{6}$ in dipole traps, in temperature,
from $\sim300\,\mathrm{K}$ in cells to $\sim30\,\mu\mathrm{K}$ in
atom traps, and in the time-scale of the interactions, from milliseconds
in cells to microseconds in cold atoms. At the same time, a variety
of other physical effects, such as loss and decoherence of atoms,
scattering and diffraction of light, and inhomogeneous coupling of
the light and atomic variables are present to varying degrees in these
systems.

We demonstrate here that these many effects can be treated within
a single framework. We work with Gaussian states, where coherent interactions\textcolor{red}{\emph{
}}and incoherent loss and decoherence processes have been studied.
Previous work on inhomogeneity has included mode matching \citep{Duan2002PRAv66p023818}
and introduction of weighted variables \citep{Kuzmich2004PRLv92p030407}.
In some cases decoherence effects due to inhomogeneous coupling have
been identified \citep{Sun2003PRAv67p063815}. Here we show how the
model of Madsen \emph{et al.} \citep{Madsen2004PRAv70p052324}, when
applied to the physical angular momentum and Stokes operators, can
be naturally extended to include inhomogeneities as well as transport
processes such as movement of atoms.

In the first part we review some important definitions of angular
momentum operators for atoms and light. We introduce the method of
segmentation and give its mathematical description in compact form.
We give a general description of physical processes including coherent
light-atom interaction, coherent and incoherent transport processes,
and projective measurements. Incoherent transport is used to describe
loss and decoherence.

In the second part we apply these techniques to calculate the effect
on spin squeezing of: imperfect detector temporal resolution, spatial
inhomogeneities in atoms and light, and atomic motion.

\section{Physical system and mathematical model}

\subsection{Continuous variables for light and atoms}

Polarized light in the framework of continuous variables can be described
in terms of the Stokes operators \begin{equation}
\hat{S}_{x}=\frac{\hbar}{2}\mathbf{a}^{\dagger}\boldsymbol{\sigma}_{x}\mathbf{a}\qquad\hat{S}_{y}=\frac{\hbar}{2}\mathbf{a}^{\dagger}\boldsymbol{\sigma}_{y}\mathbf{a}\qquad\hat{S}_{z}=\frac{\hbar}{2}\mathbf{a}^{\dagger}\boldsymbol{\sigma}_{z}\mathbf{a}\,\,.\label{eq:def_Stokes}\end{equation}
 Here $\mathbf{a}\equiv\left[\hat{a}_{+},\hat{a}_{-}\right]^{T}$
and $\hat{a}_{+},\hat{a}_{-}$ are annihilation operators for circular
plus and minus polarization, respectively and $\boldsymbol{\sigma}_{x},\,\boldsymbol{\sigma}_{y},\,\mathrm{and}\,\,\boldsymbol{\sigma}_{z}$
are the Pauli matrices. The Stokes operators have the same commutation
relations as angular momentum operators. In many situations of interest,
one polarization component is strong. Here we consider linearly polarized
light: \[
\left\langle \right.\!\hat{S}_{x}\!\left.\right\rangle =\frac{1}{2}\hbar N_{\mathrm{L}}\equiv S_{x}\qquad\mathrm{and}\qquad\left\langle \right.\!\hat{S}_{y}\!\left.\right\rangle ,\left\langle \right.\!\hat{S}_{z}\!\left.\right\rangle =0\]
 where $N_{L}$ is the number of photons. A coherent polarization
state can be expressed as an angular momentum -\textcolor{black}{\emph{
}}\textcolor{black}{\citep{Atkins1971PRSLAv321p321}}\textcolor{red}{\emph{
}}or spin and atomic - \textcolor{black}{\citep{Radcliffe1971JPAGv4p313,Arecchi1972PRAv6p2211}}
coherent state. All of them have in common that the variances orthogonal
to the main {}``spin'' are \[
\mathrm{var}(\hat{S}_{y})=\mathrm{var}(\hat{S}_{z})=\frac{1}{4}\hbar^{2}N_{\mathrm{L}}\,\,.\]
Along this direction we have either $\mathrm{var}(\hat{S}_{x})=\hbar^{2}N_{\mathrm{L}}/4$
\textcolor{black}{\citep{Atkins1971PRSLAv321p321}}\textcolor{red}{\emph{
}}or $\mathrm{var}(\hat{S}_{x})=0$ \textcolor{black}{\citep{Radcliffe1971JPAGv4p313,Arecchi1972PRAv6p2211}}
. For $N_{\mathrm{L}}\gg1$, we can substitute the operator $\hat{S}_{x}$
by its expectation value. Hence, quantum polarization features are
then solely contained in $\hat{S}_{y}$ and $\hat{S}_{z}$. Geometrically,
we are approximating a portion of the Poincaré sphere as a plane,
the geometry of the harmonic-oscillator phase space. Formally, this
is referred to as the contraction from SU(2) to the Heisenberg-Weyl
group \citep{Arecchi1972PRAv6p2211}. Consequently, the commutator
for $\hat{S_{y}}$ and $\hat{S_{z}}$ is not operator valued, as it
would be in the SU(2) algebra. Instead of writing $\left[\right.\!\hat{S}_{i},\hat{S}_{j}\!\left.\right]=\mathrm{i}\hbar\varepsilon_{ijk}\hat{S}_{k}$,
we have $\left[\right.\!\hat{S}_{y},\hat{S}_{z}\!\left.\right]=\mathrm{i}\hbar S_{x}$
and $\left[\right.\!\hat{S}_{x},\hat{S}_{y/z}\!\left.\right]=0$.
Apart from normalization, these are the commutation relations for
the generators of the Lie algebra in the Heisenberg-Weyl group.

For atoms, we similarly describe the collective spin of a collection
of atoms with the angular momentum operators \begin{equation}
\hat{J}_{x}=\frac{\hbar}{2}\mathbf{b}^{\dagger}\boldsymbol{\sigma}_{x}\mathbf{b}\qquad\hat{J}_{y}=\frac{\hbar}{2}\mathbf{b}^{\dagger}\boldsymbol{\sigma}_{y}\mathbf{b}\qquad\hat{J}_{z}=\frac{\hbar}{2}\mathbf{b}^{\dagger}\boldsymbol{\sigma}_{z}\mathbf{b}\,\,.\label{eq:def_AngMom}\end{equation}
defined in terms of bosonic operators $\mathbf{b}\equiv\left[\right.\!\hat{b}_{\uparrow},\hat{b}_{\downarrow}\!\left.\right]^{T}$.
The states $\left|\uparrow\right\rangle ,\left|\downarrow\right\rangle $
are two degenerate atomic ground states. These could be the states
of a spin-1/2 atom, as in the proposal of Kuzmich \emph{et al.} \citep{Kuzmich1998ELv42p481},
or more practically, two ground states of an alkali atom. Later, we
will consider the case of $F=1$, where $\left|\uparrow\right\rangle \equiv\left|F=1,m_{F}=-1\right\rangle ,\,\left|\downarrow\right\rangle \equiv\left|F=1,m_{F}=+1\right\rangle $
\citep{Echaniz2005JOSABv7pS548}. The operator $\mathbf{\hat{J}}$
then describes a pseudo-spin, with angular momentum commutation relations
but without spin-like behavior under spatial rotations.

We assume the atoms are polarized along a certain direction, so that
one angular momentum component can be treated classically and the
two orthogonal components carry the quantum properties. For $x$ polarization,
\[
\left\langle \right.\!\hat{J}_{x}\!\left.\right\rangle =\frac{1}{2}\hbar N_{\mathrm{A}}\equiv J_{x}\qquad\mathrm{and}\qquad\left\langle \right.\!\hat{J}_{y}\!\left.\right\rangle ,\left\langle \right.\!\hat{J}_{z}\!\left.\right\rangle =0\,\,,\]
 and the variances are\[
\mathrm{var}(\hat{J}_{y})=\mathrm{var}(\hat{J}_{z})=\frac{1}{4}\hbar^{2}N_{\mathrm{A}}\,\,.\]

\subsection{Partitioning and covariance matrix\label{sub:Partitioning}}

A central goal of this work is to include spatial inhomogeneities
in a description of the light-atom interaction. In cell experiments,
the atomic ensemble has a constant number density while a cold trapped
sample can be highly inhomogeneous. In almost all experiments, the
light distribution is inhomogeneous, e.g. from a Gaussian beam.%
\begin{figure}[h]
\includegraphics[bb=0bp 330bp 612bp 460bp,clip,width=1\columnwidth]{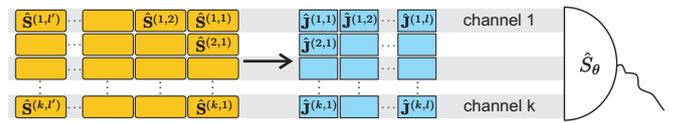}

\caption{\label{fig:dividing_scheme}Schematic of partitioned atom-light interface.
See text for details.}

\end{figure}

We split the inhomogeneous ensembles of atoms and light into several
\emph{segments}. That is, we define angular momentum variables for
the atom segments $\hat{\mathbf{J}}^{(k,l)}$, and for the light segments
$\hat{\mathbf{S}}^{(k,l')}$, with $[\hat{J}_{\lambda}^{(k,l)},\hat{J}_{\mu}^{(m,n)}]=\mathrm{i}\hbar\epsilon_{\lambda\mu\nu}\hat{J}_{\nu}^{(k,l)}\delta_{km}\delta_{ln}$
and $[\hat{S}_{\lambda}^{(k,l)},\hat{S}_{\mu}^{(m,n)}]=\mathrm{i}\hbar\epsilon_{\lambda\mu\nu}\hat{S}_{\nu}^{(k,l)}\delta_{km}\delta_{ln}$
. The segments orthogonal to the direction of light propagation are
called \emph{transverse segments} (first index) and along the direction
of light propagation, \emph{longitudinal segments} (second index)
(cf. Fig. \ref{fig:dividing_scheme}). Here we do the segmentation
in two dimensions only; the extension to the third is straightforward.
In order to stay within the assumptions of the group contraction,
we have to ensure that the particle number in each segment is itself
large, i.e., $N_{\mathrm{A}}^{(k,l)}\gg1$ and $N_{\mathrm{L}}^{(k,l')}\gg1$.
The total angular momenta for atoms and light are \begin{eqnarray}
\hat{\mathbf{J}} & = & \sum_{k,l}\hat{\mathbf{J}}^{(k,l)}\qquad\mathrm{and}\qquad\hat{\mathbf{S}}=\sum_{k,l'}\hat{\mathbf{S}}^{(k,l')}\,\,.\label{eq:splitting_J+S}\end{eqnarray}
We define a \emph{channel} as the set of segments (light and atoms)
which have the same transverse index $k$.

We assume that the ensembles both of atoms and photons can be described
as Gaussian states in the harmonic oscillator phase space and that
all operations we apply will map them into Gaussian states. For a
single segment of light we can use equation (\ref{eq:def_Stokes})
and define a phase space vector as

\begin{equation}
\hat{\mathbf{s}}^{(k,l)}\equiv\left[\!\begin{array}{c}
\hat{S}_{y}^{(k,l)}\\
\hat{S}_{z}^{(k,l)}\end{array}\!\right]=\frac{\hbar}{2}\left(\mathbf{a}^{(k,l)}\right)^{\dagger}\left\{ \!\begin{array}{c}
\boldsymbol{\sigma}_{y}\\
\boldsymbol{\sigma}_{z}\end{array}\!\right\} \mathbf{a}^{(k,l)}\,\,,\label{eq:small_s}\end{equation}
 and similarly for atoms \begin{equation}
\hat{\mathbf{j}}^{(k,l)}\equiv\left[\!\begin{array}{c}
\hat{J}_{y}^{(k,l)}\\
\hat{J}_{z}^{(k,l)}\end{array}\!\right]=\frac{\hbar}{2}\left(\mathbf{b}^{(k,l)}\right)^{\dagger}\left\{ \!\begin{array}{c}
\boldsymbol{\sigma}_{y}\\
\boldsymbol{\sigma}_{z}\end{array}\!\right\} \mathbf{b}^{(k,l)}\,\,.\label{eq:small_j}\end{equation}
 In a common phase space for atoms and light we define an overall
phase space vector in terms of angular momentum operators as \begin{equation}
\hat{\mathbf{v}}=\left[\hat{\mathbf{j}}^{(1,1)},...,\hat{\mathbf{j}}^{(k,l)},\hat{\mathbf{s}}^{(1,1)},...,\hat{\mathbf{s}}^{(k,l')}\right]^{T}\label{eq:PSV_def}\end{equation}
 which is readily rewritten as the direct sum of phase space vectors
of the sub-systems for atoms (A) and light (L) \begin{equation}
\hat{\mathbf{v}}=\hat{\mathbf{v}}_{\mathrm{A}}\oplus\hat{\mathbf{v}}_{\mathrm{L}}\label{eq:directsum_PSV}\end{equation}

Gaussian states are completely characterized by their first and second
moments. First moments $\left\langle \right.\!\mathbf{\hat{v}}\!\left.\right\rangle $
represent a displacement in phase space. Second moments or variances
are given by\begin{equation}
\boldsymbol{\gamma}=\frac{1}{2}\left\langle \mathbf{\hat{v}}\wedge\mathbf{\hat{v}}+\left(\mathbf{\hat{v}}\wedge\mathbf{\hat{v}}\right)^{\mathrm{T}}\right\rangle -\left\langle \mathbf{\hat{v}}\right\rangle \wedge\left\langle \mathbf{\hat{v}}\right\rangle \label{eq:CM_def}\end{equation}
which is the covariance matrix. For our purpose of examining entanglement
and squeezing properties, only the second moments are of interest.
We can write the covariance matrix of the joint atom-light system
as\begin{equation}
\boldsymbol{\gamma}=\left[\begin{array}{cc}
\mathbf{A} & \mathbf{C}\\
\mathbf{C}^{T} & \mathbf{L}\end{array}\right]\label{eq:AL_CM}\end{equation}
 where $\mathbf{C}$ describes correlation between atoms ($\mathbf{A}$)
and light ($\mathbf{L}$).

\section{Unified description of physical processes}

The dynamics of $\mathbf{\hat{v}}$ and $\boldsymbol{\gamma}$ is
calculated by difference equations which describe small but finite
changes between time-steps. This allows us to model coherent light-atom
interactions, losses and decoherence, measurement and transport processes
in a consistent way.

The phase space vector and covariance matrix are updated in finite
time steps $\tau$ as\begin{equation}
\hat{\mathbf{v}}(t+\tau)=\mathbf{F}_{\tau}(\hat{\mathbf{v}}(t))\,\,.\label{eq:v_trafo}\end{equation}
 The time step is chosen to be the duration of a longitudinal light
segment. It should be short enough that a longitudinal segment can
be considered homogeneous, but still contain many photons.

We note that in most experiments the atomic sample is much shorter
than the coherence time of the light pulse, which means that in each
channel only one longitudinal light segment will overlap with the
ensemble at most times. In addition, the effect of the several longitudinal
atomic segments is, from the light's perspective, sequential: $\hat{\mathbf{s}}^{(n,1)}$
interacts with $\hat{\mathbf{j}}^{(n,1)}$ then with $\hat{\mathbf{j}}^{(n,2)}$,
and so forth. In the time $t=\left(m-1\right)\tau$ to $t<m\tau$,
the $\hat{\mathbf{s}}^{(n,m)}$ interacts with all atomic segments
$\hat{\mathbf{j}}^{(n,l)}$ in the $n$-th channel.

\subsection{Coherent effects}

For effects described by a Hamiltonian $\hat{H}$ which is linear
in the elements of the phase-space vector $\hat{\mathbf{v}}$, the
phase space vector evolves as (to lowest order in $\tau$), \begin{equation}
\hat{\mathbf{v}}(t+\tau)=\hat{\mathbf{v}}(t)-\frac{i\tau}{\hbar}\left[\hat{\mathbf{v}},\hat{H}\right]\equiv\mathbf{T}_{\tau}\hat{\mathbf{v}}(t)\,\,.\end{equation}
 The last equality, which expresses the change in $\mathbf{\hat{v}}$
in terms of a matrix $\mathbf{T}_{\tau}$, is possible by the linearity
of the Hamiltonian and the c-number-valued commutation relations.
The covariance matrix evolves as \begin{equation}
\boldsymbol{\gamma}(t+\tau)=\mathbf{T}_{\tau}\boldsymbol{\gamma}(t)\mathbf{T}_{\tau}^{T}\,\,.\label{eq:int_trafo_gamma}\end{equation}

\subsubsection{Single species effects}

A magnetic field acts solely on the atomic spin and leaves the light
unchanged. Owing to the pseudo-spin character of $\hat{\mathbf{J}}$
we take only magnetic fields along the z axis into account. Such a
field results in a rotation about the z axis in the Bloch sphere.
To ensure the validity of the group contraction we also limit the
rotations to small angles. The Hamiltonian for the segment $(k,l)$
is\begin{equation}
\hat{H}_{\mathrm{magn}}^{(k,l)}=\mu_{B}g_{F}\hat{J}_{z}^{(k,l)}B_{z}^{(k,l)}\,.\label{eq:Hmagn}\end{equation}
Where, $\mu_{B}$ is the Bohr magnet on and $g_{F}$ the Landé factor.
This description includes homogeneous as well as inhomogeneous magnetic
fields.

\subsubsection{Atom-Light Interaction}

For a homogeneous system of light and atoms off-resonant interaction
gives rise to an effective Hamiltonian. For the $F=1$ pseudo-spin
system, it has the form \begin{equation}
\hat{H}_{\mathrm{eff}}\propto(\alpha^{(0)}+\frac{\alpha^{(2)}}{3})\hat{S}_{0}\hat{J}_{0}+\alpha^{(1)}\hat{S}_{z}\hat{J}_{z}+\alpha^{(2)}(\hat{S}_{x}\hat{J}_{x}+\hat{S}_{y}\hat{J}_{y})\label{eq:H_eff}\end{equation}
 where $\alpha^{(0)},\alpha^{(1)},$ and $\alpha^{(2)}$ are the scalar,
vector and tensor components of the polarizability \citep{Kupriyanov2005PRAv71p032348}.

For brevity, we will write this interaction as $\hat{H}_{\mathrm{eff}}(\mathbf{\hat{S}},\mathbf{\hat{J}})$.
As the light pulse propagates through the medium, the effects of $\hat{H}_{\mathrm{eff}}(\hat{\mathbf{s}}^{(n,1)},\hat{\mathbf{j}}^{(n,1)})$,
$\hat{H}_{\mathrm{eff}}(\hat{\mathbf{s}}^{(n,1)},\hat{\mathbf{j}}^{(n,2)})$,
and so forth are applied in sequence to the covariance matrix. Note
that loss and decoherence may be applied between these coherent evolutions.

\subsection{Noise considerations}

In addition to Hamiltonian evolution, loss, transport, and decoherence
of atoms and/or photons can be described. These processes introduce
extra noise into the system. A fully general description of a noisy
Gaussian process is the Gaussian completely-positive map (GP), which
acts on the covariance matrix as\begin{equation}
\boldsymbol{\gamma}'=\mathbf{M}\boldsymbol{\gamma}\mathbf{M}^{T}+\mathbf{N}\label{eq:GCP}\end{equation}
 where the real matrix $\mathbf{M}$ transforms the phase
space vector and the real symmetric matrix $\mathbf{N}$ describes
added noise. These must obey\textcolor{red}{\emph{ }}\textcolor{black}{\citep{Demoen1977LMPv2p161}}
\begin{equation}
\mathbf{N}+\mathrm{i}\boldsymbol{\Sigma}'-\mathrm{i}\mathbf{M}\boldsymbol{\Sigma}\mathbf{M}^{T}\geq0\label{eq:GCPNoiserequirement}\end{equation}
 where $\mathrm{i}\Sigma_{ij}\equiv[v_{i},v_{j}]$ and $\boldsymbol{\Sigma}'$,
similarly defined, are commutation matrices before and after the transformation
(note that the commutation relations, which include the \char`\"{}classical\char`\"{}
components $J_{x}$, $S_{x}$ can change due to loss and decoherence).
This places a lower limit on the noise introduced. Specifically, \begin{equation}
\mathbf{N}=|\mathrm{i}\boldsymbol{\Sigma}'-\mathrm{i}\mathbf{M}\boldsymbol{\Sigma}\mathbf{M}^{T}|\,,\label{eq:N_gen}\end{equation}
where $|\cdot|$ indicates the matrix absolute value, is the minimal
symmetric matrix to satisfy (\ref{eq:GCPNoiserequirement}).

\subsubsection{\label{sub:Loss-and-Noise}Loss and Decoherence from photon scattering}

Inevitably, the coherent interaction of equation (\ref{eq:H_eff})
will be accompanied by spontaneous emission of photons, producing
also incoherent changes in the atomic state. We use equation (\ref{eq:GCP})
to calculate the effect of loss and decoherence of atoms and photons.
Here \char`\"{}loss\char`\"{} of atoms refers to the decay of atoms
into meta-stable states which do not interact with the light. Decay
of atoms into the $\left|\uparrow\right\rangle $, $\left|\downarrow\right\rangle $
states can cause decoherence of the spin state. While loss is not
present in the ideal spin-1/2 system proposed by Kuzmich \emph{et
al.} \citep{Kuzmich1998ELv42p481}, in alkali metal atoms both processes
are observed. For light there is no decoherence process since spontaneously
emitted photons scatter into all possible spatial modes and are counted
as losses.

The covariance matrix transforms as \begin{equation}
\boldsymbol{\gamma}(t+\tau)=\mathbf{M}_{\tau}\boldsymbol{\gamma}(t)\mathbf{M}_{\tau}^{T}+\mathbf{N}_{\tau}\,,\label{eq:loss_trafo_gamma}\end{equation}
 where the decay is described by \begin{equation}
\mathbf{M}_{\tau}=(1-\eta_{\tau})\mathbb{I}_{2}\oplus(1-\varepsilon)\mathbb{I}_{2}\,.\label{eq:D}\end{equation}
Here $\eta_{\tau}$ and $\varepsilon$ are scattering probabilities
for an atom and a photon, respectively. For rubidium 87 these are
given in terms of experimental parameters in the appendix. Noise will
have the form $\mathbf{N}_{\tau}=\mathbf{N}_{\tau,\mathrm{\mathrm{l}oss}}+\mathbf{N}_{\mathrm{\tau,\mathrm{d}ec}}$
with (\ref{eq:N_gen}) we get \begin{eqnarray}
\mathbf{N}_{\mathrm{\tau,\mathrm{l}oss}} & = & \eta_{\tau}(1-\eta_{\tau})\frac{\hbar^{2}}{4}N_{\mathrm{A}}\mathbb{I}_{2}\oplus\varepsilon(1-\varepsilon)\frac{\hbar^{2}}{4}N_{\mathrm{L}}\mathbb{I}_{2}\,\,.\label{eq:noise_from_loss}\end{eqnarray}
 and \begin{eqnarray}
\mathbf{N}_{\mathrm{\tau,\mathrm{d}ec}} & = & \rho\eta_{\tau}\frac{\hbar^{2}}{4}N_{\mathrm{A}}\mathbb{I}_{2}\oplus\mathbb{O}_{2}\,\,.\label{eq:noise_from_dec}\end{eqnarray}
Here $\rho$ is the fraction of the scattered atoms which return to
the system, assumed to be in a mixed state. This model has been used
in the literature \citep{Hammerer2004PRAv70p044304} and serves to
illustrate the method. A different model would be necessary to describe
some processes, e.g., optical pumping. $\mathbb{I}_{\mathrm{2}}$
is the identity matrix in two dimensions. $\mathbb{O}_{\mathrm{2}}$
is the zero matrix and reflects the fact that we don't consider any
decoherence for the light. For all simulations that follow in section
\ref{sec:Results} we assume we have exclusively atomic decoherence
and no loss, i.e., $\rho=1$.

\subsection{\label{sub:Measurement}Projective Measurement}

The next class of operations we can apply are measurements of atomic
or light variables. While a measurement will collapse the value of
an observable in a way that is fundamentally random, the resulting
variances change in a way that is completely predictable: The variance
of the measured observable becomes zero, the variance of the conjugate
observable becomes large or infinite. The variances of other observables
may also be reduced if they are correlated with the measured observable.

A measurement can be described by a projection matrix $\mathcal{P}$.
For example, to measure a polarization component of the $(n,i)$ light
segment, $\hat{S}_{\theta}^{(n,i)}\equiv\cos\theta\hat{S}_{y}^{(n,i)}+\sin\theta\hat{S}_{z}^{(n,i)}\equiv(\mathbf{p}_{\theta}^{(n,i)})^{T}\cdot\mathbf{v}$,
the projector would be the outer product $\mathcal{P}=\mathbf{p}_{\theta}^{(n,i)}\wedge\mathbf{p}_{\theta}^{(n,i)}$.
In practical situations, measurement of a light variable also implies
that a light segment has reached a detector and thus is removed from
the problem, reducing the dimension of the vector $\mathbf{v}$. Upon
measurement, the covariance matrix becomes \begin{equation}
\boldsymbol{\gamma}'=\left|\boldsymbol{\gamma}-\boldsymbol{\gamma}(\mathcal{P}\boldsymbol{\gamma}\mathcal{P})^{-}\boldsymbol{\gamma}^{T}\right|_{(n,i)}\,\,.\label{eq:CM_meas_update}\end{equation}
 Where $\left|...\right|_{(n,i)}$ removes the column and row corresponding
to the measured, and no longer existing, light segment $(n,i)$. $(...)^{-}$
indicates the Moore-Penrose pseudoinverse. Equation (\ref{eq:CM_meas_update})
is well known in mathematical statistics to compute the conditional
covariance matrix of multivariate normal distributions \citep{Marsaglia1964JotASAv59p1203}.
A more detailed introduction of Gaussian operations on Gaussian states
can be found in \citep{Eisert2002PRLv89p137903,Giedke2002PRAv66p032316}.

For the calculations in Part 2, we consider a large-area detector,
i.e., one which does not distinguish between different channels (see
Fig. \ref{fig:dividing_scheme}). Therefore, we define the measured
light variable to be \begin{equation}
\hat{S}_{\theta}^{(l)}={\displaystyle \sum_{k=1}}\hat{S}_{\theta}^{(k,l)}\,\,.\label{eq:large_det}\end{equation}

\subsection{Combining effects}

When several effects are present at the same time-step, they are applied
sequentially. The order of application can influence the results of
the calculation if the time-step is not small. For example, when the
light-atom interaction of Eq. (\ref{eq:int_trafo_gamma}) and noise
of Eq. (\ref{eq:loss_trafo_gamma}) are both considered, we can have
\begin{equation}
\boldsymbol{\gamma}(t+\tau)=\mathbf{M}_{\tau}\mathbf{T}_{\tau}\boldsymbol{\gamma}(t)\mathbf{T}_{\tau}^{T}\mathbf{M}_{\tau}+\mathbf{N}_{\tau}\label{eq:noise_last}\end{equation}
 or \begin{equation}
\boldsymbol{\gamma}(t+\tau)=\mathbf{T}_{\tau}\left[\mathbf{M}_{\tau}\boldsymbol{\gamma}(t)\mathbf{M}_{\tau}+\mathbf{N}_{\tau}\right]\mathbf{T}_{\tau}^{T}\,\,,\label{eq:noise_first}\end{equation}
 depending on which effect is applied first. Physically, this ordering
has no meaning, and in the limit of small time steps $\tau$, both
(\ref{eq:noise_last}) and (\ref{eq:noise_first}) give the same result.
In the simulations that follow, we reduce $\tau$ until the effect
of the ordering is negligible.

\section{Results\label{sec:Results}}

Now we give three examples how the model can be applied in the context
of atomic spin squeezing. For all simulations we consider a cold ensemble
of rubidium 87 atoms in a dipole trap. The set of used parameters
can be found in the appendix. It is well known that for large detunings
from resonance we can reduce the dipole interaction Hamiltonian (\ref{eq:H_eff})
to

\begin{equation}
\hat{H}_{\mathrm{eff}}\left(\mathbf{\hat{s}}^{(k,j)},\mathbf{\hat{j}}^{(k,l)}\right)=\frac{\hbar g^{(k,l)}}{\tau}\hat{S}_{z}^{(k,j)}\hat{J}_{z}^{(k,l)}\,\,.\label{eq:H_QND}\end{equation}
The coupling constant $g^{(k,l)}$ is proportional to the vector polarizability
$\alpha^{(1)}$ and defined in the appendix.  Note, the Hamiltonian
does not explicitly depend on $\tau$ because the Stokes operators
are proportional to the flux of photons times $\tau$. As initial
states we assume a pulse of horizontally polarized light, i.e., $S_{x}=N_{\mathrm{L}}\hbar/2$
and a coherent superposition of the Zeeman substates $\left|F=1,m_{F}=-1\right\rangle $
and $\left|F=1,m_{F}=1\right\rangle $, i.e., $J_{x}=N_{\mathrm{A}}\hbar/2$.

The effect of the Hamiltonian (\ref{eq:H_QND}) on the light is a
rotation of $\hat{S}_{x}$ about the $z$-axis by an amount proportional
to $\hat{J}_{z}$. At the same time, $\hat{J}_{z}$ is not altered
in this process. A projective measurement of $\hat{S}_{y}$ provides
information about, and thus reduces the uncertainty of, $\hat{J}_{z}$.
If both input states are minimum uncertainty states, spin squeezing
is obtained. To monitor the evolution of this process we evaluate
$2\mathrm{Var}\left(\right.\!\hat{J}_{z}\!\left.\right)/J_{x}$ which
is also known as the spin squeezing parameter \citep{Kitagawa1993PRAv47p5138}.
For squeezed states it will become less than unity. The smaller the
spin squeezing parameter the higher the degree of spin squeezing.
There are other criteria, for instance by Wineland \emph{et al.} \citep{Wineland1992PRAv46pR6797}
derived in the context of precision spectroscopy. Regardless which
of the definitions is applied, we obtain the same qualitative results.

To make the comparison between different experimental situations clearer,
we normalize the timescale. We can define a time when the rotation
of the light polarization due to the atom-light interaction (\ref{eq:H_QND})
exceeds the shot noise of the photons, i.e., when the signal-to-noise
ratio becomes one. We want this time to be characteristic for the
system as a whole. Therefore, we neither partition atoms nor light
and get \begin{equation}
t_{0}=\frac{4}{\hbar^{4}}\frac{1}{G^{2}N_{\mathrm{A}}\Phi}\,\,.\label{eq:typ_time}\end{equation}
Where $G$ is the collective interaction strength, $N_{\mathrm{A}}$
the number of atoms and $\Phi$ the photon-flux. A detailed derivation
is given in the appendix.

\subsection{Detector time-resolution}

As a first example, we study the influence of the detector time-resolution
on the amount of spin squeezing and show the importance of correct
modeling of pulsed experiments even for pulses much shorter than the
detector time resolution.

We define an \emph{ideal detector} as one capable of detecting individual
light segments. The covariance matrix would be updated in accordance
to (\ref{eq:CM_meas_update}) each time a light segment hits the detector.
In contrast, we say a detector has no time-resolution if it detects
all segments at the same time. Mathematically, the measured variable
is the sum of all $n$ light segments\begin{equation}
\hat{S}_{y}^{(1)}={\displaystyle \sum_{l=1}^{n}}\hat{S}_{y}^{(1,l)}\,\,,\label{eq:S_y}\end{equation}
and we apply the transformation (\ref{eq:CM_meas_update}) to the
whole covariance matrix. (For simplicity we assume only one atomic
segment. Nonetheless, we keep the transverse index to avoid confusion.)
The projector has the form \begin{equation}
\mathcal{P}=\mathbb{O}_{2}\oplus\,\frac{1}{n}\mathbb{U}_{n}\otimes\mathbb{I}_{2}\,\,.\label{eq:P}\end{equation}
 Where $\mathbb{U}_{n}$ is the unit matrix of rank $n$.

\begin{figure}[!h]
\begin{centering}
\includegraphics[bb=0bp 150bp 612bp 650bp,clip,width=1\columnwidth]{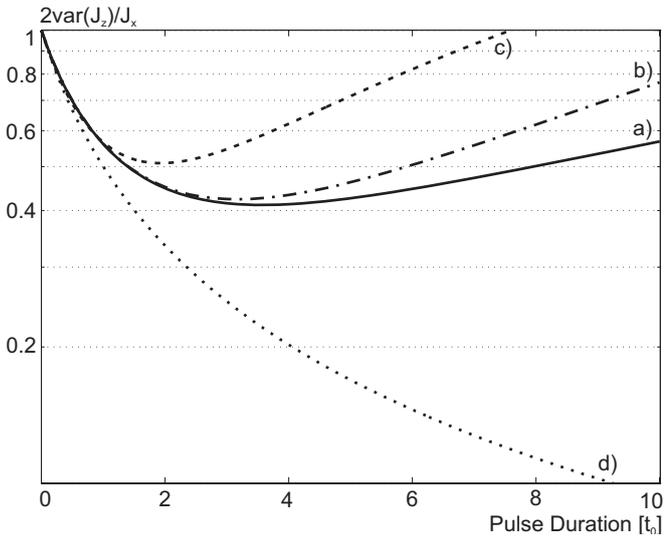}
\par\end{centering}

\caption{\label{fig:ITR}The spin squeezing achieved by using an ideal detector
(a) and a detector with no temporal resolution (b) are shown. The
parameters are given in appendix A. Two zero-dimensional calculations
are given for comparison. In curve (c) the noise due to decoherence
and loss was added after and in (d) before the interaction. }

\end{figure}

In Fig. \ref{fig:ITR} we show the results for both having (a) perfect
and (b) no temporal resolution. In the case of no temporal resolution,
the achievable spin squeezing is reduced at longer timescales. This
is understood if we compare the information carried by different longitudinal
light segments. Early light segments interact with the initial atomic
state and later ones with a noisier version of it. If the detector
is lacking temporal resolution all this different information is mixed.

Now we compare the results to calculations which neglect all dynamics
during the pulse duration, e.g., in \citep{Kuzmich2004PRLv92p030407}.
We call this type of model {}``zero-dimensional'' because it treats
the light-atom interaction as a point-like event in time. Therefore,
we assume that the light pulse is not partitioned into longitudinal
segments. Curve (c) and (d) in Fig. \ref{fig:ITR} show the results
if the noise is added after (cf. Eq. \ref{eq:noise_last}) and before
(cf. Eq. \ref{eq:noise_first}) the interaction, respectively. It
becomes obvious that even in the case of a pulsed experiment it is
important to model light as a stream of sufficiently short segments.%

\subsection{Spatial inhomogeneities}

In many experiments inhomogeneities in light or atomic distributions
are present. We give two examples for typical situations that can
arise. As the simplest test model we assume an atomic ensemble which
consists of two equally sized transverse segments.

For all the following calculations we assume an ideal detector. Furthermore,
the total number of atoms $N_{A}$, the photon flux $\Phi$, the total
interaction cross section $A$, and all other parameters are fixed
and stated in the appendix. The figure of merit is the variance of
$\hat{J}_{z}$ for the complete atomic ensemble\begin{equation}
\mathrm{var}\left(\hat{J}_{z}\right)=\mathrm{var}\left(\sum_{i}\hat{J}_{z}^{(i)}\right)\,\,.\label{eq:}\end{equation}

To verify the validity of the segmentation model, we consider first
a homogeneous atom distribution either as a single or as two segments.
In both cases we observe the same results in the presence of atom-light
interaction, loss and decoherence and transport processes, independent
on the segmentation. They reproduce the solid line (a) in Fig. \ref{fig:ITR}).

\begin{figure}[h]
\begin{centering}
\includegraphics[bb=10bp 0bp 770bp 612bp,clip,width=1\columnwidth]{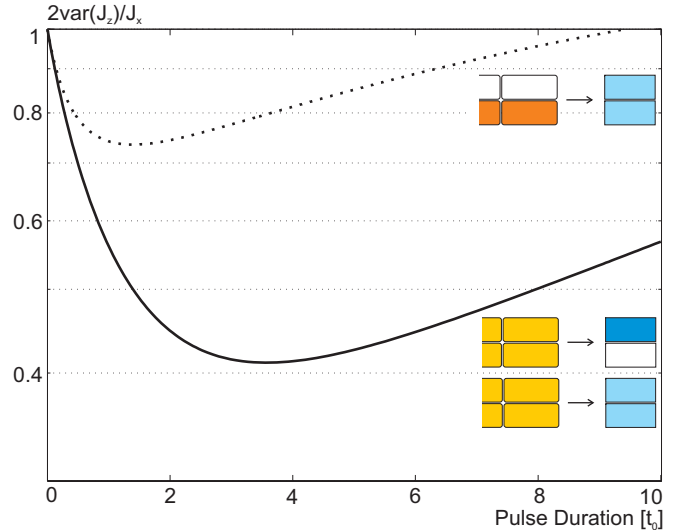}
\par\end{centering}

\caption{\label{fig:spat_inhomo}Two exemplary cases are compared to the homogeneous
situation were light and atoms are evenly distributed over space.
For more information see text.}

\end{figure}

The first example reflects the situation we would find for inhomogeneous
light fields interacting with homogeneously distributed atoms. We
model this with two channels of equal interaction cross-section $A/2$.
We assume light is only present in one of the channels. The result
is plotted as the dotted curve in Fig. \ref{fig:spat_inhomo}. The
overall spin squeezing is reduced. To explain this, we can evaluate
both channels independently. One channel contains all the photons
and the maximal obtainable amount of squeezing will be the same as
for the homogeneous distribution (solid line in Fig. \ref{fig:ITR}).
This reflects a very important property in atomic spin squeezing.
The achievable amount of squeezing does not depend on the intensity
of light (supposed it is not zero) but rather on the optical depth
of the atomic ensemble. For the second channel, without light, we
expect no change in the atomic state. If we combine these two results,
we get exactly the dotted curve shown in Fig. \ref{fig:spat_inhomo}.

The second example is the inverse situation. The light beam has a
larger cross-section than the atomic ensemble. We model this case
by assuming all $N_{A}$ atoms only in one of the channels and light
homogeneously distributed over both. The result, plotted as the solid
line in Fig. \ref{fig:spat_inhomo}, seems surprising. We see the
exact same dynamics as for the homogeneous case. In this situation
two effects are compensating each other. The optical depth for the
atoms is twice as large as in the previous examples and leads to larger
spin squeezing. On the other hand, the light which does not interact
with the atoms is also detected, and contributes noise but no additional
information about the atoms.

The two examples give some intuition about the influence of inhomogeneities.
It is now straightforward to apply it to more interesting and complicated
experimental cases.

\subsection{Atomic motion}

As a last application, we ask what happens when atoms can change places
and go from one segment to another. This is a relevant question comparing
different experimental situations. Atoms in vapor cells, for instance,
have approximately room temperature. This corresponds to a root-mean-square
(rms) velocity of hundreds of meters per second as opposed to a few
tens of millimeters per second for dipole-trapped atoms. Typical timescales,
$t_{0}$, for the light-atom interaction are milliseconds and microseconds,
respectively. Hence, atoms in vapor cells have moved around half a
meter (effectively) whereas the atoms in the trap moved less than
hundred nanometers. This suggests that for trapped atoms, any inhomogeneity
in the light beam will be mapped onto them %
\footnote{For atoms in buffer gas vapor cells one can expect a similar behavior.%
}. To find a more quantitative description, we introduce a mixing probability
$m_{\tau}$ per time step $\tau$ and per atom. It is defined as the
probability an atom would escape from one segment to another in one
time step.

As in the previous section, we use the simple test model of two channels,
where all the light is concentrated in one part and the atoms are
homogeneously distributed over both. The real symmetric matrix describing
the mixing is

\begin{eqnarray}
\mathbf{M} & = & \left[\!\begin{array}{cc}
1-m_{\tau} & m_{\tau}\\
m_{\tau} & 1-m_{\tau}\end{array}\!\right]\otimes\mathbb{I}_{2}\,,\label{eq:M}\end{eqnarray}
and the introduced noise is given by

\begin{equation}
\mathbf{N}=m_{\tau}\left(1-m_{\tau}\right)\frac{\hbar^{2}}{4}N_{\mathrm{A}}\left[\!\begin{array}{cc}
1 & -1\\
-1 & 1\end{array}\!\right]\otimes\mathbb{I}_{2}\,.\label{eq:N}\end{equation}
\begin{figure*}[!t]
\begin{centering}
\includegraphics[bb=0bp 170bp 792bp 450bp,clip,width=1\textwidth]{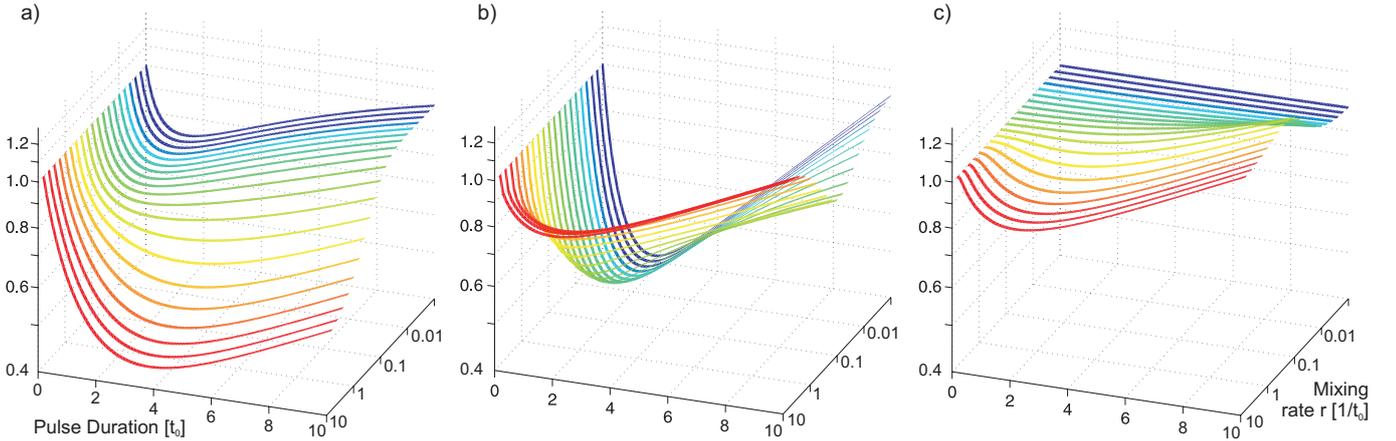}
\par\end{centering}

\caption{\label{fig:diff_tc} Atomic motion is simulated with different mixing
rates. a) shows the degree of spin squeezing for the whole ensemble,
b) for the illuminated and c) for the un-illuminated segment. The
discussion is given in the text.}

\end{figure*}
The noise matrix $\mathbf{N}$ reflects two things. First, the individual
variance in a single segment will increase. This is not surprising,
because already correlated atoms leave the segment and uncorrelated
atoms enter. Second, the variances of the total spin components are
not altered by mixing. This is expected, since the choice of the partitioning
is arbitrary and can have therefore no influence. An alternate way
to derive $\mathbf{N}$ and $\mathbf{M}$ is sketched in %
\footnote{Interestingly, one can arrive at the same results for $\mathbf{N}$
and $\mathbf{M}$ assuming a unitary matrix which connects the bosonic
atom(light)-modes of different segments. The mixing matrix is then
defined as $M_{ij}=\left|U_{ij}\right|^{2}$. Whereas, the noise matrix
drops out automatically, when we allow for arbitrary phases between
the different input modes and average over them. Which is equivalent
to say that atomic mixing is incoherent. %
}

As a concrete example we assume we have an ideal gas of atoms and
derive the mixing probability from kinetic  gas theory. The number
of collisions per area and time in an ideal gas is known to be $R=Nv_{\mathrm{rms}}/\left(V\,\sqrt{6\pi}\right)$.
Where, $N$ is the number of atoms in volume $V$, and their rms velocity
is $v_{\mathrm{rms}}$. From this we can calculate the rate at which
individual atoms cross a surface of area $A'$, $r=RA'/N$. Furthermore,
we assume that the atoms occupy a box of volume $V=A'\,\sqrt{A}$,
where $A$ is the interaction cross section of a segment. The rate
can therefore be written as\begin{equation}
r=\frac{v_{\mathrm{rms}}}{\sqrt{6\pi A}}\,\,.\label{eq:r_final}\end{equation}
In the limit of small $\tau$ we can define a mixing probability as
$m_{\tau}=r\tau$.

\begin{equation}
m_{\tau}=\frac{v_{\mathrm{rms}}}{\sqrt{6\pi A}}\tau\,\,.\label{eq:m_tau}\end{equation}
Where $v_{\mathrm{rms}}=\sqrt{3k_{B}T/m}$ is the rms velocity of
the atoms.\emph{ }This model is not an exact treatment of the different
physical situations we find in vapor cells and atomic traps. Nevertheless,
it suggests how atomic motion influences the formation of spin squeezing.

In Fig. \ref{fig:diff_tc} we plot the squeezing factor for different
mixing probabilities. For $m_{\tau}/\tau\rightarrow0$ (blue curves)
we have the same situation as in the dotted curve of Fig. \ref{fig:spat_inhomo}.
For increasing mixing probability (red curves) we see that the squeezing
is improved and the full amount (compared to the homogeneous situation,
i.e., solid curve of Fig. \ref{fig:ITR}) is achieved again. In this
limit, when the inverse mixing rate becomes the same order of magnitude
as $t_{0}$, sufficient atomic movement is present that all the atoms
get enough interaction to be uniformly squeezed. This suggest that
approaches of matched variables, e.g., by Kuzmich \emph{et al.} \citep{Kuzmich2004PRLv92p030407}
are more relevant for cold atoms than for hot vapor cell experiments.

If we instead focus our attention only on the segment of atoms which
is illuminated we see in plot b) of Fig. \ref{fig:diff_tc} that the
spin squeezing for this segment reduces when the mixing probability
increases. One can interpret this as a decoherence mechanism for the
smaller segment \citep{Sherson2006IB}.

\section{Conclusion}

We have presented a model to compute the dynamics of interacting light
and atomic ensembles with Gaussian states. The model is based on covariance
matrices for the quantum components of collective angular momentum
operators and employs segmentation of the light and atom systems.
The model is similar to that of Madsen \emph{et al.} \citep{Madsen2004PRAv70p052324},
but extends the segmentation to light and to the transverse directions.
Also, we use angular momentum operators, rather than derived canonical
operators, which give intuitive results and simplify partitioning.
We show how to include many effects which arise in real experimental
situations, including spatial and temporal inhomogeneities, atom motion,
loss, and noise introduced by photon scattering.

Employing this model, we have made the following observations: The
dynamics of spin squeezing requires time-dependent modeling, even
when the atoms interact with optical pulses which are shorter than
the detection system can resolve. At the same time, the detector time
resolution has only a minor effect on the degree of spin squeezing
under realistic conditions. The effect of spatial inhomogeneities
in light and atoms have non-equivalent effects on spin squeezing:
Concentration of the light into a sub-region of the atoms produces
equal squeezing of the sub-region, but less squeezing of the entire
ensemble, while concentration of the atoms into one sub-region of
the light gives equal squeezing of the spin ensemble. Finally, we
observe that atomic motion between an illuminated region and a non-illuminated
one tends to degrade squeezing of the illuminated region while increasing
squeezing of the entire atomic ensemble. This suggests that high-fidelity
experiments should use probe pulses which are either much shorter,
or much longer, than the time-scale of the atomic motion.

The model can straight-forwardly be adapted to more complicated experimental
situations, for example a cold thermal cloud in a focused laser beam.
Also, application to multi-pass schemes as proposed by by Takeuchi
\emph{et al}. \citep{Takeuchi2005PRLv94p023003} or Sherson \emph{et
al.} \citep{Sherson2006PRAv74p011802} is possible.

\begin{acknowledgments}
We thank O. S. Mishina, M. Lewenstein, M. Kubasik, S. R. de
Echaniz, and M. Napolitano for helpful discussions. This work was
funded by the Spanish Ministry of Science and Education under the
LACSMY project (Ref. FIS2004-05830) and the Consolider-Ingenio
2010 Project \textquotedblleft{}QOIT\textquotedblright{}.
\end{acknowledgments}

\section*{Appendix}

The atomic ensemble we consider has $10^{6}$ rubidium 87 atoms at
a temperature of $30\,\mu\mathrm{K}$. For the light we assume a flux
$\Phi=10^{14}\, s^{-1}$ of linearly polarized photons with a detuning
$\Delta=1\,\mathrm{GH}\mathrm{z}$ from the $F=1\rightarrow F'=0$
transition of the $D_{2}$ line. The corresponding wavelength in vacuum
is $\lambda=780.241\,\mathrm{nm}$. Both atoms and photons interact
over a cross section of $A=4\,\pi\times10^{-10}\,\mathrm{m}^{2}$.

The coupling constant $g$ in (\ref{eq:H_QND}) is directly related
to the vector part of the polarizability tensor \citep{Geremia2006PRAv73p042112}.
For the $F=1$ hyperfine ground state in the limit of detunings larger
than the natural linewidth we find\begin{equation}
g^{(k,l)}=\frac{1}{A^{(k,l)}}\frac{\Gamma\lambda^{2}}{16\pi}\frac{1}{\hbar^{2}}\left(-4\delta_{0}\!\left(\Delta\right)-5\delta_{1}\!\left(\Delta\right)+5\delta_{2}\!\left(\Delta\right)\right).\label{eq:def_couplconst}\end{equation}
Where $A^{(k,l)}$ is the interaction cross-section of the segment
$(k,l)$ and $\Gamma/2\pi=6.065\,\mathrm{MHz}$ is the natural line
width of the $5P_{3/2}$ excited state. The functions $\delta_{F'}(\Delta)=\left(\Delta+\Delta_{0,F'}\right)^{-1}$
include the finite hyperfine splittings in the excited state: $\Delta_{0,F'}$
is the hyperfine level spacing between $F'=0$ and $F'=1,2$. It would
be straightforward to include inhomogeneity in the local light shift,
due to the dipole trap, as $\Delta^{(k,l)}$. However, we don\textasciiacute{}t
consider it for the simulations given here.

The scattering probabilities for photons and atoms are given by $\eta_{\tau}=N_{\mathrm{L,\tau}}\sigma(\Delta)/A$
and $\varepsilon=N_{\mathrm{A}}\sigma(\Delta)/A$, as in \citep{Hammerer2004PRAv70p044304}.
Where, $A$ is the interaction area and $N_{\mathrm{L},\tau}$ and
$N_{\mathrm{A}}$ the number of photons and atoms in a segment, respectively.
The off-resonant atomic scattering cross-section is \begin{equation}
\sigma\left(\Delta\right)=\frac{\lambda^{2}}{2\pi}\frac{\Gamma^{2}}{32}\left(4\delta_{0}\!\left(\Delta\right)^{2}+5\delta_{1}\!\left(\Delta\right)^{2}+7\delta_{2}\!\left(\Delta\right)^{2}\right)\,\,,\label{eq:sigma_D}\end{equation}
which is valid for detunings much larger than the natural line width.

We derive the characteristic time $t_{0}$. If we apply the interaction
(\ref{eq:H_QND}) for a time $\tau$ we find \begin{eqnarray}
\hat{S}'_{y} & = & \hat{S}_{y}+G\hbar\hat{J}_{z}S_{x}\nonumber \\
\hat{J}'_{y} & = & \hat{J}_{y}+G\hbar\hat{S}_{z}J_{x}\,\,,\label{eq:whole_system}\end{eqnarray}
where $G=\frac{1}{A}\frac{\Gamma\lambda^{2}}{16\pi}\frac{1}{\hbar^{2}}\left(-4\delta_{0}\!\left(\Delta\right)-5\delta_{1}\!\left(\Delta\right)+5\delta_{2}\!\left(\Delta\right)\right)$
similar to (\ref{eq:def_couplconst}). The variances are readily calculated
for coherent input states (unprimed) \begin{eqnarray}
\mathrm{var}(\hat{S}'_{y}) & = & \frac{\hbar^{2}}{4}N_{\mathrm{L,t_{0}}}+G^{2}\hbar^{2}\frac{\hbar^{2}}{4}N_{\mathrm{A}}\frac{\hbar^{2}}{4}N_{\mathrm{L,t_{0}}}^{2}\nonumber \\
 & = & \frac{\hbar^{2}}{4}N_{\mathrm{L,t_{0}}}\left(1+G^{2}\hbar^{2}\frac{\hbar^{2}}{4}N_{\mathrm{A}}N_{\mathrm{L,t_{0}}}\right)\label{eq:}\end{eqnarray}
The same also holds for $\mathrm{var}(\hat{J}'_{y})$. When the second
term in brackets is unity this describes a signal-to-noise ratio of
one and with $N_{\mathrm{L,\tau}}=\Phi\tau$ this occurs for $\tau=t_{0}$
as given in (\ref{eq:typ_time}). If we use the parameters given above
$t_{0}$ is $0.55\,\mu\mathrm{s}$.

\bibliography{CM_paper_8.bbl}

\end{document}